\documentclass[ApJL,twocolumn]{aastex631}

\DeclareUnicodeCharacter{2212}{-}
\usepackage{booktabs}	
\usepackage{color, colortbl}
\usepackage{multirow}

\begin{document}

\title{Imaging and Spectral Fitting of Bright Gamma-ray Sources with the COSI Balloon Payload}

\author[0000-0002-7660-2740]{Jarred M. Roberts}
\affiliation{University of California, San Diego,
Dept. of Astronomy and Astrophysics,
9500 Gilman Dr., La Jolla, CA 92093, USA}

\author[0000-0001-9567-4224]{Steven Boggs}
\affiliation{University of California, San Diego,
Dept. of Astronomy and Astrophysics,
9500 Gilman Dr., La Jolla, CA 92093, USA}

\author[0000-0002-0552-3535]{Thomas Siegert}
\affiliation{Julius-Maximilians-Universität Würzburg, Fakultät für Physik und Astronomie, Institut für Theoretische Physik und Astrophysik,\\
Lehrstuhl für Astronomie, Emil-Fischer-Str. 31, D-97074 Würzburg, Germany}

\author[0000-0001-5506-9855]{John A. Tomsick}
\affiliation{University of California, Berkeley,
Space Sciences Laboratory, 
7 Gauss Way, Berkeley, CA 94720}

\author[0000-0002-6584-1703]{Marco Ajello}
\affiliation{Clemson University, 
Department of Physics and Astronomy,
118 Kinard Laboratory, Clemson, SC 29634}

\author[0000-0002-0917-3392]{Peter von Ballmoos}
\affiliation{Institut de Recherche en Astrophysique et Planétologie, Université de Toulouse, CNRS, CNES, Université Toulouse III Paul Sabatier, 9 avenue Colonel Roche, BP 44346, 31028 Toulouse Cedex 4, France}

\author[0000-0002-0128-7424]{Jacqueline Beechert}
\affiliation{University of California, Berkeley,
Space Sciences Laboratory, 
7 Gauss Way, Berkeley, CA 94720}

\author{Floriane Cangemi}
\affiliation{Universit\'{e}-Paris Cit\'{e},
45 Rue des Saints-P\`{e}res, 75006 Paris, France}


\author[0000-0002-2664-8804]{Savitri Gallego}
\affiliation{Institut für Physik \& Exzellenzcluster PRISMA+,Johannes Gutenberg-Universität Mainz, 55099 Mainz, Germany}

\author[0000-0002-1757-9560]{Pierre Jean}
\affiliation{Institut de Recherche en Astrophysique et Planétologie, Université de Toulouse, CNRS, CNES, Université Toulouse III Paul Sabatier, 9 avenue Colonel Roche, BP 44346, 31028 Toulouse Cedex 4, France}

\author[0000-0002-6774-3111]{Chris Karwin}
\affiliation{NASA Goddard Space Flight Center,
8800 Greenbelt Rd, Greenbelt, MD 20771}

\author[0000-0001-6677-914X]{Carolyn Kierans}
\affiliation{NASA Goddard Space Flight Center,
8800 Greenbelt Rd, Greenbelt, MD 20771}

\author{Hadar Lazar}
\affiliation{University of California, Berkeley,
Space Sciences Laboratory, 
7 Gauss Way, Berkeley, CA 94720}

\author{Alex Lowell}
\affiliation{University of California, Berkeley,
Space Sciences Laboratory, 
7 Gauss Way, Berkeley, CA 94720}

\author[0000-0002-2471-8696]{Israel Martinez Castellanos}
\affiliation{NASA Goddard Space Flight Center,
8800 Greenbelt Rd, Greenbelt, MD 20771}
\affiliation{Center for Research and Exploration in Space Science and Technology, NASA/GSFC, Greenbelt, MD 20771}
\affiliation{Department of Astronomy, University of Maryland, College Park, MD 20742}

\author[0000-0002-8403-0041]{Sean Pike}
\affiliation{University of California, San Diego,
Dept. of Astronomy and Astrophysics,
9500 Gilman Dr., La Jolla, CA 92093, USA}

\author[0000-0003-4732-6174]{Clio Sleator}
\affiliation{Naval Research Laboratory,
4555 Overlook Ave SW, Washington, DC 20375}

\author[0000-0002-3833-1054]{Yong Sheng}
\affiliation{Clemson University, 
Department of Physics and Astronomy,
118 Kinard Laboratory, Clemson, SC 29634}

\author[0000-0002-5345-5485]{Hiroki Yoneda}
\affiliation{Julius-Maximilians-Universität Würzburg, Fakultät für Physik und Astronomie, Institut für Theoretische Physik und Astrophysik,\\
Lehrstuhl für Astronomie, Emil-Fischer-Str. 31, D-97074 Würzburg, Germany}

\author[0000-0001-9067-3150]{Andreas Zoglauer}
\affiliation{University of California, Berkeley,
Space Sciences Laboratory,
7 Gauss Way, Berkeley, CA 94720}

\begin{abstract}

The Compton Spectrometer and Imager balloon payload (COSI-Balloon) is a wide-field-of-view Compton $\gamma$\,-ray telescope that operates in the 0.2\,--\,5\,MeV bandpass.
COSI-Balloon had a successful 46-day flight in 2016 during which the instrument observed the Crab Nebula, Cygnus\,X-1, and Centaurus\,A.
%
%
%
Using the data collected by the COSI-Balloon instrument during this flight, we present the source flux extraction of signals from the variable balloon background environment and produce images of these background-dominated sources by performing Richardson-Lucy deconvolutions.
We also present the spectra measured by the COSI-Balloon instrument, compare and combine them with measurements from other instruments, and fit the data.
The Crab Nebula was observed by COSI-Balloon and we obtain a measured flux in the energy band 325\,--\,480\,keV of (4.5\,$\pm$\,1.6)\,$\times$10$^{−3}$\,ph\,cm$^{-2}$\,s$^{-1}$.
The model that best fits the COSI-Balloon data combined with measurements from NuSTAR and \textit{Swift}-BAT is a broken power law with a measured photon index $\Gamma = 2.20\,\pm\,0.02$ above the 43\,keV break. 
Cygnus\,X-1 was also observed during this flight, and we obtain a measured flux of (1.4\,$\pm$\,0.2)\,$\times 10^{−3}\,$ph\,cm$^{-2}$\,s$^{-1}$ in the same energy band and a best-fit result (including data from NuSTAR, \textit{Swift}-BAT, and INTEGRAL/IBIS) was to a cutoff power law with a high-energy cutoff energy of $138.3\,\pm\,1.0\,$keV and a photon index of $\Gamma = 1.358\,\pm\,0.002$.
Lastly, we present the measured spectrum of Centaurus A and our best model fit to a power law with a photon index of $\Gamma = 1.73\,\pm\,0.01$.
%
%

\end{abstract}


\section{Introduction}
Three of the brightest sources in the MeV $\gamma$-ray sky are the Crab Nebula (Crab), Cygnus\,X-1 (Cyg\,X-1), and Centaurus\,A (Cen\,A). 
Although these sources have been studied extensively across a wide band of energies, there are fewer observations in the MeV bandpass.
The region of the bandpass from 100\,keV to 100\,MeV is commonly referred to as the \textit{MeV Gap}, due to the lack of sensitivity in this band \citep{Siegert2022_gammaraytelescopes}.
The Compton Spectrometer and Imager (COSI) is a NASA-confirmed $\gamma$-ray space explorer designed to probe the MeV Gap, currently being built with an anticipated deployment date of late 2027 \citep{Tomsick_ICRC_2023}.
We present images and spectral results obtained from the balloon-borne predecessor to the space instrument (COSI-Balloon) using analysis tools currently under development ahead of the primary space mission.
%


\begin{figure}[!t]
    \centering
    \includegraphics[width=0.46\textwidth,trim=0.0in 0.0in 0.0in 0.0in,clip=True]{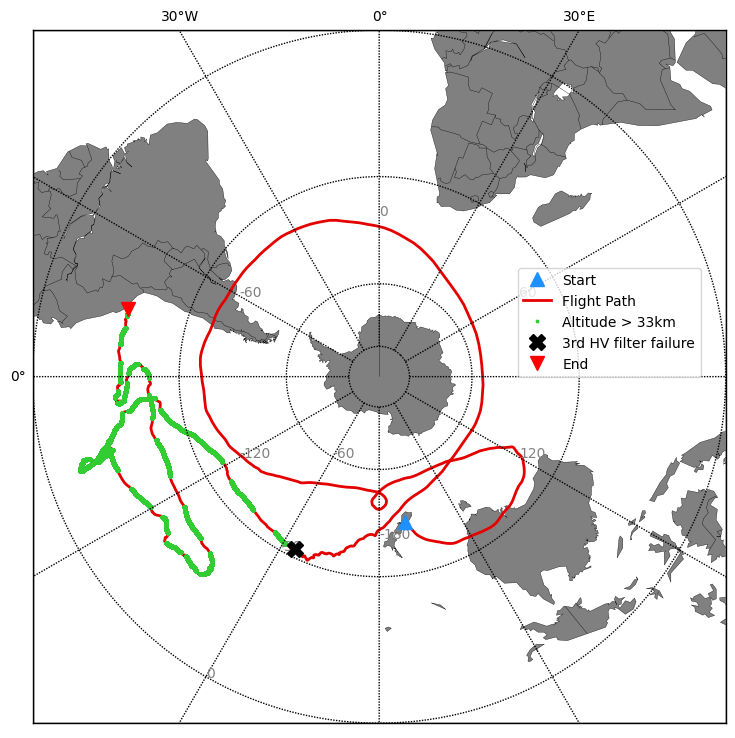}
    \caption{\footnotesize 
    The 46-day super-pressure balloon flight path of the 2016 COSI-Balloon instrument (red trace). The flight path in green represents the portions of the flight used throughout this analysis.}
    \label{fig:balloon_path}
\end{figure}


\subsection{Crab}
The Crab Nebula is the remnant of a core-collapse supernova that exploded ${\sim}$1000 years ago. The Crab is one of the most studied objects in the $\gamma$-ray sky and is commonly used by high-energy instruments as a calibration source.
For example, it has been observed by CGRO-COMPTEL \citep[0.75\,--\,30\,MeV][]{COMPTEL_Crab_1997,COMPTEL_Crab_1995},
INTEGRAL/SPI \citep[20\,keV\,--\,8\,MeV][]{SPI_Crab_2009,Jourdain2020_Crab}, INTEGRAL/IBIS \citep[15\,keV\,--\,10\,MeV][]{Jourdain_Crab_2008,IBIS_Crab_2010}, \textit{Swift}-BAT \citep[15\,--\,150\,keV][]{BAT_Crab_2010} and also at higher energies by Fermi-LAT \citep[20\,MeV to $>$ 300\,GeV][]{LAT_Crab_2010,LAT_Crab_2011}. 
Although the Crab spectrum is consistent with a single power law in the X-ray band \citep{Madsen_2015}, this has been seen to be softer in the MeV band.
The spectral shape of the Crab has been described by a Band model between 20\,keV and 2.2\,MeV \citep{Jourdain2020_Crab}.
The averaged spectrum best-fit parameters in \cite{Jourdain2020_Crab} corresponded to a low-energy slope of $1.99 \pm 0.01$, and a high-energy slope of $2.32\,\pm\,0.02$.
The spectral shape using the instruments onboard INTEGRAL (JEM-X, ISGRI, and SPI) can be described by a broken power law with a photon index of  $2.105\,\pm\,0.003$ below the break energy and a photon index of $2.22\,\pm\,0.02$ above the break energy \citep{Jourdain_Crab_2008}. 
The spectrum measured by the COSI-Balloon instrument is discussed in Sec.\,\ref{sec:Crab_results}.

\subsection{Cygnus X-1}
Cyg\,X-1 is a binary star system that is comprised of a black hole and a massive giant companion star \citep{Cygx1_bh_1972,Orosz_Cygx1_2011}.
Cyg\,X-1 is one of the most studied X-ray binary systems due to its bright and persistent X-ray and $\gamma$-ray emission. 
The two main canonical states are the low-hard state (LHS) and the high-soft state (HSS) \citep{belloni2010_cygx1_states}. 
In the LHS, the X-ray spectrum between ${\sim}$1\,keV and ${\sim}$\,100\,keV is dominated by a hard power law with a photon index of $\Gamma \leqslant 1.5$ with a high-energy cutoff at a few hundred keV \citep{belloni2010_cygx1_states}.
Compact radio jets are only observed when an X-ray binary source is in the LHS and has been resolved in Cyg\,X-1 \citep{Fender_cygx1_jet_2001,Sterling_cygx1_jet_2001}

\par
In the HSS, the spectrum is dominated by a multi-temperature blackbody component peaking at ${\sim}$\,1\,keV and is associated with thermal emission from an optically thick accretion disk.
No radio emission is typically detected when sources are in the HSS state \citep{Fender_cygx1_1999,Remillard_cygx1_2006,belloni2010_cygx1_states} with the possible exception of Cyg\,X-1 where there has been some evidence for radio emission being detected using very long baseline interferometry \citep{Rushton_cygx1_2012}.

\par
Long-term variability studies performed by RXTE have found that the most stable state of Cyg\,X-1 is the LHS, followed by the HSS \citep{Grindberg_cygx1_states_2013}.
The probability that the source remains in the LHS for at least one week (200\,h) is $>85\%$.
Soft states are less stable, but the probability of a soft state lasting longer than one week is still ${\sim}$\,75\%. 
Intermediate states are short-lived and typically last about 24 hours.
The existence of a high-energy tail extending above a few hundred keV is of particular interest for Cyg\,X-1 because it suggests that particles are accelerated in the vicinity of the black hole, \textit{e.g.}, the corona and jet.
\par
The COMPTEL experiment on CGRO measured evidence for a high-energy tail above several hundred keV while the source was observed in the LHS \citep{McConnell_2000}.
A power law best described the spectrum at MeV energies with a photon index of $\Gamma = 2.58\,\pm\,0.03$ above 750\,keV and no evidence of a high-energy cutoff at high energies \citep{McConnell_2000}.
Observations of Cyg\,X-1 in the HSS by CGRO, where the spectrum is dominated by the emission of non-thermal electrons and Compton reflection in the CGRO bandpass \citep[$\geq$\,10\,keV][]{gierlinski_cygx1_1999}, led to a model fit consistent with a power-law-like emission extending up to 10\,MeV with no high-energy cutoff and an increase in flux above 1\,MeV in comparison to the LHS \citep{McConnell_2002}. 
\par
Long-term studies performed by INTEGRAL (JEM-X, IBIS, and SPI) show a high-energy tail while the source is in both the LHS and HSS \citep{Cangemi2021_CygX1_hardtail}.
This measurement builds on the existing evidence of the presence of a high-energy component of the source, which extends to at least 600\,keV, which falls within the bandpass of the COSI-Balloon $\gamma$-ray instrument.
Results from the COSI-Balloon observations of Cyg\,X-1 can be found in Sec.\,\ref{sec:CygX1_results}.
%

\begin{table}[t]
\centering
\caption{\footnotesize Event selections for Crab, Cyg\,X-1, and Cen\,A}
\begin{tabular}{| c | c |}
\hline
 Parameter                        & Selection                             \\ 
 \hline
 \hline
 Number of gamma-ray interactions &  2--7                                 \\
 \hline
 Distance between first           &  $>0.5$\,cm                           \\
 two interactions                 &                                       \\
 \hline 
 Distance between other           &  $>0.3$\,cm                           \\
 interactions $(3+)$              &                                       \\
 \hline
 Energy Selections                & $\gtrsim40$\,keV -- $\lesssim5$\,MeV  \\
 \hline
 Compton scatter angle            & $[0,60^{\circ}]$                      \\
 \hline
 Altitude                         &  $>33$\,km                            \\
 \hline
 Pointing Coordinates             &  Galactic (full-sky)                  \\
 \hline
 Disk radius around known         &  $[0,60^{\circ}]$                     \\
 source coordinates               &                                       \\
 \hline
 Earth Horizon Cut                &     Applied                           \\ 
 \hline 
\end{tabular}
\label{table:1}
\end{table}


\subsection{Centaurus A}
Cen\,A (NGC\,5128) is one of the brightest radio-loud active galactic nuclei (AGNs) in the hard X-ray and soft $\gamma$-ray sky. 
Cen\,A has been observed by numerous missions since its first high-energy detection in 1969 \citep{Bowyer_cena_1970}.
At a few tens of keV to GeV photon energies, Cen\,A was detected by all instruments on board the CGRO (BATSE, OSSE, COMPTEL, and EGRET), revealing a high-energy peak in the spectral energy density (SED) at an energy of ${\sim}0.1$\,MeV \citep{Kinzer_cena_1995,Steinle_cena_comptel_1998,Sreekumar_cena_EGRET_1999}.
The discovery of Cen\,A as an emitter of very high energy (VHE) $\gamma$-rays was reported by the High Energy Stereoscopic System (H.E.S.S.) from observations performed from April 2004 to July 2008 \citep{Reiger_HESS_cena_2009}.
This VHE signal from the region containing the radio core, the parsec-scale jet, and the kiloparsec-scale jet was detected with a statistical significance of 5\,$\sigma$.
Subsequent survey observations at high energies (HE; 100\,MeV\,--\,100\,GeV) were performed by the LAT onboard Fermi \citep{Atwood_FERMI}. 
During the first three months of science operation, Fermi-LAT confirmed the EGRET detection of the Cen\,A $\gamma$-ray core \citep{Abdo_FERMI_cena_2009}. 

\par
Various observations have indicated that the spectral shape of Cen\,A does not vary with flux \citep{Jourdain_cena_1993,Rothschild_2006,Rothschild_2011,Beckmann_cena_2011,Burke_2014}.
Although it is expected that the reported spectral parameters for Cen\,A will be similar across time and instruments, measurements have shown variability.
There has been a lack of consistency between values, which has made interpreting the origin of the hard X-ray/soft $\gamma$-ray emission challenging.

\par
The presence of spectral curvature and where the break in the spectra occurs have significant implications for determining the emission process. 
Hard X-ray continuum in AGNs is widely interpreted as thermal inverse Comptonization of UV seed photons from the accretion disk \citep{Haardt_1993,Merloni_2003}.
The spectra typically have emissions up to a few hundred keV \citep{Molina_2013}, and the spectral cutoff energy is related to the temperature of electrons in the corona.
In the corona, photon-photon collisions can create electron-positron pairs when the sum of the photon energies exceeds the pair-creation threshold \citep{Svensson_1984,Stern_1995,Fabian_AGN_2015}.
If the electron temperature is too high, the production of electron-positron pairs can become a runaway process that reduces the temperature of the corona.

\par
An analysis of Cen\,A using INTEGRAL/ISGRI and SPI covering roughly 20 years of observations found no significant deviations from
a constant power law \citep{Rodi_2023}.
Joint NuSTAR/ISGRI/SPI spectral fit from this analysis supported a cutoff energy of ${\sim}650$\,keV or an electron temperature of ${\sim}$\,520\,--\,550\,keV.
A log-parabola also accurately describes the spectrum to model synchrotron self-Compton emission from the jet. 
When the Fermi-LAT spectrum in the 50\,MeV\,--\,100\,GeV range was included in the spectral fit, a log-parabola model for the jet emission explains the spectrum from NuSTAR to LAT below ${\sim}$\,3\,GeV. 
Flux from the jet provided the best explanation for the model fits from X-rays to the GeV region with one component sufficient to explain the whole emission from keV to GeV and the high electron temperature implied by the corona scenario. 
The spectrum measured by the COSI-Balloon observation of Cen\,A is presented in Sec.\,\ref{sec:CenA_results}, and the implications of the model fit are discussed.

\section{The Compton Spectrometer and Imager} \label{sec:intro}
COSI-Balloon was the balloon-payload predecessor to the Small Explorer space instrument, COSI \citep{Tomsick_ICRC_2023,Tomsick2019_COSI}, planned for launch in 2027.
The COSI-Balloon instrument was a soft $\gamma$-ray (0.2\,--\,5\,MeV) telescope designed to be deployed on a NASA 18 million cubic-foot
Super-Pressure Balloon (SPB). 
The detector head of COSI-Balloon was composed of twelve high-purity germanium (HPGe) cross-strip detectors \citep{Amman_2007}.
Each detector is approximately 8\,cm\,$\times$\,8\,cm\,$\times$\,1.5\,cm in volume, coated with amorphous germanium \citep{Amman_2018}, and fabricated with 37 aluminum strip electrodes deposited orthogonally on opposite sides of each detector.
With a strip pitch of 2\,mm, the internal position sensitivity results in a 3D voxel size of about 1.5\,mm$^3$ \citep{lowell_2016}.
These HPGe detectors provided optimal spectral resolution of about $0.6\%$ FWHM at 662\,keV \citep{BEECHERT2022166510}.
The twelve detectors were stacked in a $2\,\times\,2\, \times\,3$ configuration in an aluminum cryostat with a mechanical cooler and a total active volume of 972\,cm$^3$. 
The cryostat was surrounded by cesium iodide (CsI) scintillator anti-coincidence shields on four outward-facing sides and bottom (Earth-facing) to act as an active background radiation veto.
The COSI-Balloon cryostat was fixed on top of a zenith-oriented gondola frame and operated as a free-floating, non-pointing, wide field-of-view surveying instrument.
%


\begin{figure}[!t]
    \centering
    \includegraphics[width=0.45\textwidth,trim=0.0in 0.0in 0.0in 0.0in,clip=True]{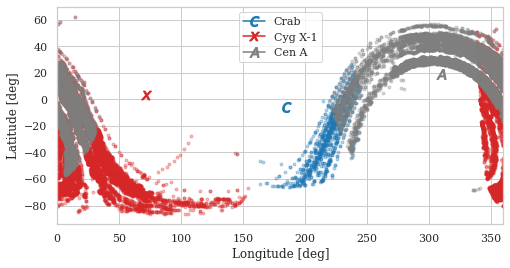}
    \caption{\footnotesize Instrument pointings throughout the 2016 balloon flight using data selections shown in Tab.\,\ref{table:1}. The labels $C$, $X$, and $A$ represent the Galactic source locations for the three sources (Crab, Cyg\,X-1, and Cen\,A), respectively. The points represent the zenith pointings of the COSI instrument as per the data selections for each data set.}
    \label{fig:pointings}
\end{figure}


\subsection{The 2016 COSI-Balloon Flight}
The COSI-Balloon instrument had a successful 46-day flight on board a SPB in 2016 \citep{Kierans2017:COSIBalloon}.
The payload launched from Wanaka, New Zealand, on May 17\textsuperscript{th} and was terminated 200\,km northwest of Arequipa, Peru, on July 2\textsuperscript{nd}, where it was successfully recovered.
The nominal flight altitude was 33\,km, with some altitude variations due to day-night cycles. 
During the balloon flight, three high-voltage filters failed, preventing three detectors from receiving the required high-voltage bias and making them inactive.
Two of these detectors were located at the top of the 4 by 3 detector stack, and the 3\textsuperscript{rd} detector was located in the second/middle row.
The result of these inactive detectors was a reduction of the instrument's sensitivity by ${\sim}40\%$ around 500\,keV \citep{Siegert2020:511}. 

After launch, the circumpolar winds carried the payload around the Southern Ocean, circling the continent of Antarctica for about 14\,days.
The balloon then drifted to more northern latitudes and meandered around the equatorial Pacific Ocean for the remainder of the flight.
Finally, the payload was terminated from the balloon, the parachute deployed, and the payload landed on the west coast of Peru. 
\par
The path of the balloon during this flight is shown in Figure\,\ref{fig:balloon_path}.
During the first half of the flight, the anti-coincidence shields measured high background rates that had no strong correlation with the flight altitude.
The data from this first half of the flight are excluded from these analyses because developing a background model would be prohibitively complex.
The red trace in Figure\,\ref{fig:balloon_path} indicates times/regions where the data are excluded from this analysis, and the green traces indicate the portions of the flight where the data passes our selections for quality.
Also illustrated in this figure, during the second half of the flight, the balloon was floating at lower latitudes (\textit{i.e.}, closer to the equator), which influences the geomagnetic cut-off rigidity and, consequently, the background rate.
After the 29\textsuperscript{th} day, the balloon and payload experienced frequent altitude changes, which increased the background count rates during drops in altitude. 
These strong correlations will be used to determine the variation of the instrumental background and determine appropriate background tracers to fit the data.
Applying the optimal data selections and background models was critical for the analysis of the $\gamma$-ray source addressed in this work. 
\par
Analyses already produced from this flight data range from the first $\gamma$-ray burst (GRB) reported to the GRB Coordination Network from a balloon-borne $\gamma$-ray instrument \citep{Tomsick_GRB160530A_2016}, $\gamma$-ray polarization \citep{Lowell2017_COSI_polarisation}, 511\,keV spectral analysis \citep{Kierans2018_PhD}, 511\,keV image analysis \citep{Siegert2020:511}, $^{26}\mathrm{Al}$ spectral analysis \citep{Beechert_2022}, and Galactic diffuse emission \citep{Karwin_2023}. 
The scope of this work will be to present the first images of three point-like sources (Crab, Cyg\,X-1, and Cen\,A) using a Richardson-Lucy maximum likelihood technique and spectral analyses of all three sources, using data collected by the COSI-Balloon $\gamma$-ray telescope.  
%


\begin{figure}[!t]
    \centering
    \includegraphics[width=0.48\textwidth,trim=0.0in 0.0in 0.0in 0.0in,clip=True]{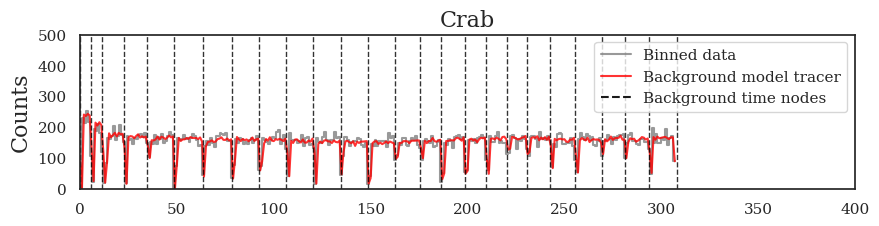}
    \includegraphics[width=0.48\textwidth,trim=0.0in 0.0in 0.0in 0.0in,clip=True]{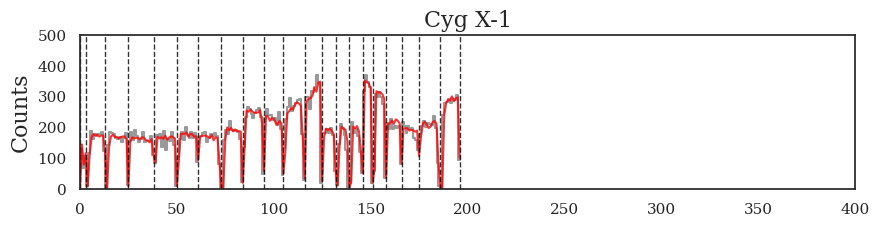}
    \includegraphics[width=0.48\textwidth,trim=0.0in 0.0in 0.0in 0.0in,clip=True]{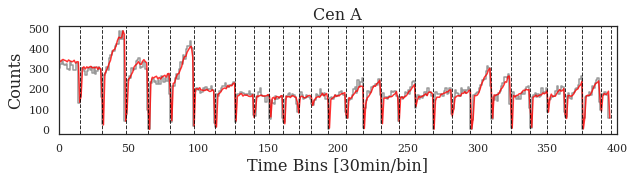}
    \caption{\footnotesize Crab, Cyg\,X-1, and Cen\,A data selections binned into 30\,min time bins (gray). A background-fitted tracer is then applied to the data (red), and background time \textit{nodes} are then
    applied to the data to account for the constantly changing background (vertical black dashed lines). Note that the total number of counts in each time bin, which results in drops when the pointing cut is applied. These then serve as natural time nodes for re-scaling the background.}
    \label{fig:background_bins}
\end{figure}


\pagebreak
\section{The COSI-Balloon Data Space} \label{sec:Data}

The COSI-Balloon instrument records individual events within the active detector volume, and the event reconstruction is performed using the deposited energy and the kinematics of Compton scattering (\textit{e.g}, \cite{Zoglauer2007_EventReconstruction,vonBallmoos1989_ComptonTelescope,Boggs2000_EventReconstruction,2021arXiv210213158Z}). 
The stored parameters are the total photon energy $E$ per event, the three scattering angles in the Compton Data Space: $\phi$ (Compton scattering angle; $[0,180^{\circ}]$), $\psi$ and $\chi$ (the angles in spherical coordinate encoding the direction of the scattered gamma, $[0,180^{\circ}]$, $[−180^{\circ},180^{\circ}]$ ), and the event time. 
The COSI-Balloon data space consists of a tag for the energy and time of each event in the 3-dimensional {$\phi\,\psi\,\chi$} data space; This is commonly referred to as the COMPTEL (or Compton) data space (CDS) \citep{schoenfelder_1999_CDS}.
The aspect of the COSI-Balloon instrument is also saved as both the pointing of the detector head in $x$ and $z$ optical axis in both galactic and horizon coordinate systems. 

\subsection{Data Selections and Processing}

All event data from the balloon flight are first processed through calibration and event reconstruction software included in the Medium-Energy Gamma-ray Astronomy library (MEGAlib, \cite{Zoglauer2006_MEGAlib}).
We perform a variety of event selections, starting with the exclusion of all data collected before the failure of the third high-voltage filter. 
During the majority of this early section of the flight, the Crab and Cyg\,X-1 are below the Earth's horizon.
The exclusion of this data does not significantly impact the observation time of these sources.
Drops in altitude that occurred during the second half of the flight resulted in higher background count rates that are difficult to fit. 
Our event selections exclude all data taken when the balloon and instrument dropped below 33\,km.

\par
We also perform a data selection of the calibrated and event-reconstructed data sets, which follow a more detailed quality control of the individual Compton events.
For these individual events, we only select those with kinematic Compton reconstruction chain lengths (\textit{i.e.}, photon interactions) of 2 to 7. 
Events with three or more interactions within the detector(s) lead to a higher fraction of correctly reconstructed events \citep{Zoglauer2006_MEGAlib}, and removing events with two interactions would reduce the instrument's sensitivity. 
Reconstructions also become increasingly computationally expensive for multi-scattered events that follow $N!$ permutations ($8! =  40320$ possible paths).
Therefore, we restrict the total number of interactions for an event to $\leq\,7$.
\par
COSI-Balloon's angular resolution is dependent on the position resolution determined by the 2\,mm strip pitch of the GeDs.
A minimum distance between the first and second Compton interactions of 0.3\,cm was chosen to optimize the angular resolution at the expense of detection efficiency. 
Applying selections to the Compton scattering angle can also improve the quality of the data set because back scatters, where the photon scatters at an angle $>90^{\circ}$, are difficult to reconstruct.
A disk radius selection ($\phi$) around the source locations is used during event extractions to better isolate the source depending on the observation, the position of the source, and the locations of other known luminous sources within the field-of-view (FOV). 
This disk radius selection acts as a timing cut, accepting all interactions of any reconstructed origin within the detector that pass all other quality selections, but only while the source is within a selected FOV.
These selections are described as \textit{pointing cuts}.
The specific event selections used to produce the images for all three sources presented in this analysis are summarized in Tab.\,\ref{table:1}. 
The data selections used to generate the spectral fits are similar, except that pointing cuts of $90^{\circ}$ are used. 

\pagebreak
\subsection{Data Binning}

For the Compton space, \textit{e.g.} with bin sizes of $1^{\circ}$ in the CDS, this would correspond to $180^{\circ} \times 180^{\circ} \times 360^{\circ}$ = 11,664,000 bins, and would result in a number of data points that would be difficult to process computationally.
For this reason, we set the angular binning to $6^{\circ}$ as this is comparable to COSI-Balloon's angular resolution in the energy band where COSI-Balloon has maximal sensitivity (325\,--\,480\,keV).
The spatial resolution capabilities of Compton-style telescopes are energy-dependent, position-dependent (at higher incident angles of a point source, the angular resolution decreases), and are defined by the angular resolution measure (ARM).
For COSI-Balloon, the ARM of a 356 keV photon with an off-axis angle of 30$^{\circ}$ from zenith is ${\sim}8^{\circ}$\,FWHM \citep{SLEATOR2019162643, BEECHERT2022166510}.
The ARM of lower energy photons (${\sim}$\,300\,keV) of a source with off-axis angles of ${\sim}30^{\circ}$ is ${\sim}9^{\circ}$\,FWHM.
With $6^{\circ}$ angular binning of the images produced in this work and discussed in Sec. \ref{sec:imaging}, we expect the majority of the source flux to fall within pixels within the ARM radius for the given energy bin.
 
\par
With COSI-Balloon being a wide FOV, zenith-pointing instrument suspended below the SPB, the gondola's position (aspect) is constantly changing. 
The data must then be divided into time bins of the appropriate length in order to account for the changing position and background. 
The time bins used for these analyses should not be too long because different exposures with and without signals will be combined together in time.
They should also not be too short, as the limited number of counted photons would lead to an unnecessarily large data space.
For these analyses, we adopt a time binning of 30 minutes and weigh different impacts on the imaging and background response accordingly within each time bin.
We further reduce the number of bins in the data space by identifying CDS bins that are never occupied (\textit{i.e.} in the response, data, and background).
We remove these bins, which is similar to adapting a sparse representation of the data space. 
These empty data bins can exist in the data set, the background, and also in the imaging response. 

\section{Data Analysis}

 We model the COSI-Balloon data, $d$, as a linear combination of a sky model, $s$, and a background model, $b$, with unknown amplitudes $\alpha$ and $\beta$, respectively. 
 The data are divided into 10 energy bins, with bin edges in keV: 150,  220,  325,  480,  520,  765, 1120, 1650, 2350, 3450, and 5000.
 The model used to describe the COSI-Balloon data is represented by the equation:
\begin{equation}
m_i=\alpha s_i + \beta b_i
\label{equ:model}
\end{equation}
The following sections describe model templates $s$ and $b$ in further detail.
Being that photon counting is a Poisson process, the likelihood that data $d$ is produced by a model $m$ is described by the Poisson likelihood:
\begin{equation}
\mathcal{L}(d|m) = \prod_{i=1}^{N} \frac{m_i^{d_i}e^{-m_i}}{d_i!}
\label{equ:poisson}
\end{equation}
where $N$ is the number of CDS bins per energy bin, $m$ is the sky model, and $d$ is the number of observed counts in each data space bin.
This function can be more simply understood as the likelihood that a given physical model $m$ accurately describes the observed data, $d$, which is equal to the probability of obtaining the particular observed data sample, $d$, in the CDS bin, $i$.
The physical sky model is composed of source parameters considered for the analysis, \textit{e.g.}, the flux of the sources, the spectral index of the sources, their location in the sky, background rates, etc., and the observed data corresponds to the measured counts in each bin in the 3-dimensional Compton data space, \{$\phi\,\psi\,\chi$\}.  

We fit for the scaling factors $\alpha$ and $\beta$ by minimizing the Cash statistic \citep{Cash1979_cstat}, which is the negative logarithm of the likelihood in the Poisson distribution (\ref{equ:poisson}) being agnostic to model-independent terms:
\begin{equation}
\mathcal{C}(d|m) = -\sum_{i=1}^{N} [m_i - d_i \ln(m_i)]
\label{equ:cash}
\end{equation}
In contrast to the imaging response, which is derived from simulations, the background model is determined empirically and will be
treated as being unitless. 
More detailed descriptions of the data analysis methods for the COSI-Balloon instrument are presented in \cite{Siegert2020:511}.

\subsection{Instrument Imaging Response}\label{sec:image_resp}

The imaging response maps the image space to the CDS described in Sec.\,\ref{sec:Data}. 
Assuming a binned image space spanning the celestial sphere in Galactic coordinates, the imaging response encodes the effective area and describes the probability that an event emitted in the given image bin (with the given energy and possible polarization) is detected in a given data space bin.
We use the Medium-Energy Gamma-ray Astronomy library \citep[MEGAlib][]{Zoglauer2006_MEGAlib} to simulate the expected number of photons within each energy bin as a function of the intrinsic zenith and azimuth coordinate system, for the full 2016 balloon flight.
This simulation requires a detailed mass model of the COSI-Balloon instrument that needs to account for the three inactive detectors and the attenuation of the atmosphere at specific altitudes.
The instrument's effective area is a strong function of atmospheric attenuation, as encoded through the zenith angle; incident photons must pass through more air mass at the same zenith angle when the instrument is at lower altitudes. 
In fact, the point spread function also changes due to atmospheric scattering in a way that \textit{more} photons can be detected than without the atmosphere and is described in detail in \cite{Karwin:2024kcl}.

\par
The simulation setup uses a mass model of the COSI-Balloon detector head with 9 active detectors in the center of an isotropically symmetric sphere with the instrument at an altitude of 33\,km.
The simulated events are then processed by a well-benchmarked detector effects engine that converts the simulated events into data that mimics what would be collected by the COSI-Balloon instrument (\textit{e.g.} detector strip numbers and ADC values, \cite{SLEATOR2019162643}).
The simulated events are then further processed with the same calibration pipeline used for real data \citep{Beechert_2022}.
After the events are reconstructed, they are binned according to a pre-defined spacing in the 5-dimensional data space that is defined by the zenith and azimuthal angles in detector coordinates (Z,A), as well as the Compton data space, \{$\phi\,\psi\,\xi$\}.
Here, we use a $6^\circ$ binning for all angles, resulting in 54,000 CDS bins and 1,800 sky pixels. 
The sparse CDS is then typically populated by less than 10\%.
%


\begin{figure*}[ht]
    \centering
    \includegraphics[width=0.52\textwidth,trim=0.0in 0.0in 0.0in 0.0in,clip=True]{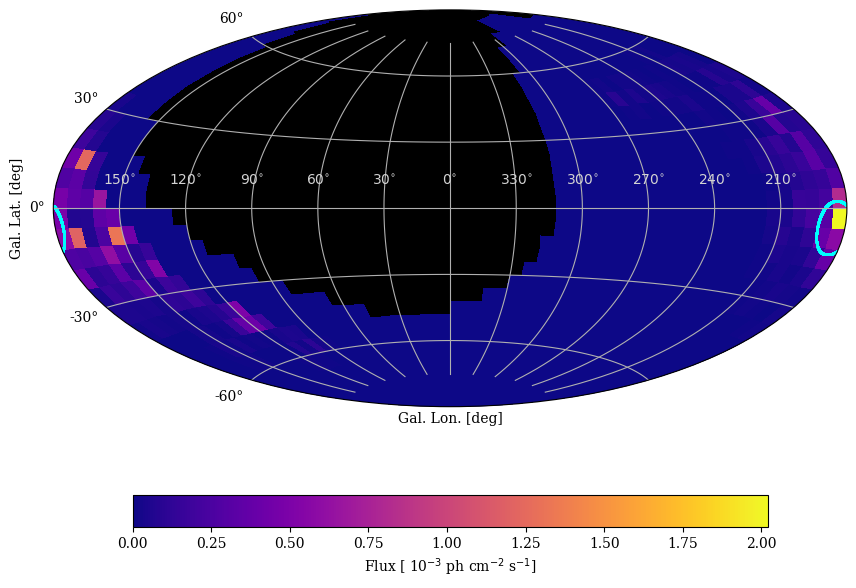}
    \includegraphics[width=0.47\textwidth,trim=0.0in 0.0in 0.0in 0.0in,clip=True]{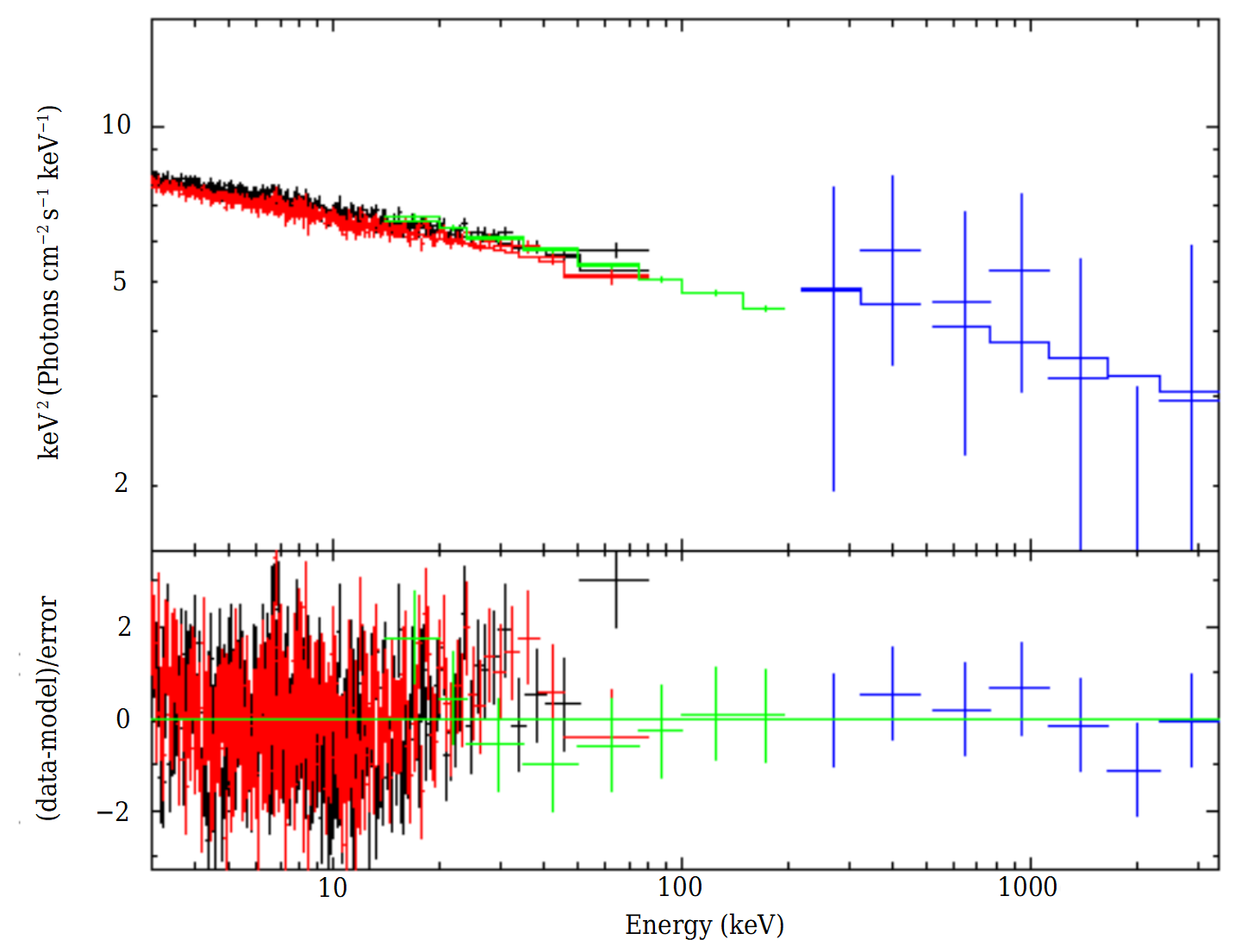}
    \caption{\footnotesize \textbf{Left:} Crab image generated from the 2016 COSI-Balloon flight data. The pixels are separated into $6^{\circ}$ bins, and the cyan circle drawn at the known source location with an ARM radius of 8\,$^{\circ}$ is shown to represent the instrument angular resolution for the energy bin 325\,--\,480\,keV.  \textbf{Right:} Crab spectrum as measured by NuSTAR (black, red), Swift/BAT (green), and the COSI-Balloon instrument (blue), fit with a broken power-law. The fit has a reduced-$\chi ^2$\ value of 1.08 (see Tab.\,\ref{tab:Crab})}
    \label{fig:CRAB_Swift_BAT_spec}
\end{figure*}


\subsection{Background Modeling}
Astronomical observations in the low-energy $\gamma$-ray regime are dominated by background; A signal-to-noise ratio of 1:100 or worse is not uncommon \citep{Lubinski:2008zt}.
An estimated expected background spectrum and its intensity can be created with computationally intensive simulations that account for the full mass model of the instrument payload and precise environmental conditions.
A significant source of instrumental background for space instruments and high-altitude balloon-borne instruments comes from the interaction of primary cosmic-ray particles with the payload materials \citep{Cumani_2019, Boggs2002_SPIBG, Gehrels1985_balloonBG}.
Secondary particles in the MeV energy range lead to nuclear excitations followed by de-excitations with the emission of $\gamma$-rays with short and long-lived radioactive nuclei.

\par
There is also a dominant instrumental continuum background created by $\beta$-particles depositing energy within the detector volume and also electromagnetic cascades from high-energy cosmic rays.
The cosmic ray flux is heavily dependent on the solar cycle, and the additional unpredictable occurrences of solar flares restrict the feasibility of a physical background model that can be applied to every time and position.
The $\gamma$-ray instrumental background of balloon environments is complicated further by atmospheric effects.
Cosmic-ray particles interacting in the upper atmosphere create a strong, soft $\gamma$-ray continuum \citep{Cumani2019_MeVBG}.
Secondary electrons also produce bremsstrahlung that results in down-scattered $\gamma$-ray photons that can add to the instrumental background.
\par
The low-energy background below 10\,MeV at scientific balloon altitudes described by \cite{Ling1975_MeVBG} can be used to perform simulations to assess the adequacy of background inference and was used to generate the background models in this analysis.
We build an empirical background model with two components (the shape of the background in the CDS and a tracer function that determines the variation of the background in time) that are parameterized in variability and amplitude and divide the data into background time nodes that sub-divide the tracer function into a pre-defined number of subsets with amplitudes for each time interval \citep{Siegert2020:511}.
This model is then fitted simultaneously along iterative image deconvolution to describe the celestial emission.
These tracers are then fitted simultaneously with the sky model amplitude, $\alpha$ (Eq.\,\ref{equ:model}).
Figure\,\ref{fig:background_bins} shows the time-binned data sets for the Crab, Cyg\,X-1, and Cen\,A, the model tracers applied to each data set, and the background time nodes chosen for each data set used to generate the images.
Background time nodes are not applied to the data sets used to generate the individual spectra of these sources.
\par
Atmospheric scattering correction factors have been applied to the spectral data as described in \cite{Karwin:2024kcl}, using a simplistic rectangular atmospheric model and the average off-axis angle for each source.
We estimate that the effects from atmospheric scattering from a point source would reduce the total photon flux by ${\lesssim}10\%$ \cite{Karwin:2024kcl}.
Given the large uncertainties in the fitted spectra, the flux reduction due to atmospheric scattering is not statistically significant in this work.
We nevertheless apply the atmospheric correction factors to all three spectra presented in Sections, \ref{sec:Crab_results}, \ref{sec:CygX1_results}, and \ref{sec:CenA_results}.
All data processing and spectral fitting of the COSI-Balloon data is performed using the initial version of \texttt{COSIpy} upon which also the 511\,keV imaging analysis was performed \citep{Siegert2020:511}, while the model fitting of the spectra that include multiple data sets is performed using \texttt{Xspec} \citep{xspec_1999}.
%


\begin{table}[ht]
\centering
\caption{Spectral Results for Crab Broken Power Law Fit\label{tab:Crab}}
\begin{tabular}{| c | c | c | }
\hline
\hline
 Parameter\footnote{The errors on the parameters are 90\% confidence.}      &Unit            &Value\\ 
 \hline
 \hline
 \multicolumn{3}{|c|}{Tied Parameters\footnote{The full XSPEC model is: constant*TBabs*bknpower}} \\
 \hline
 $N_H$          & $10^{22}$\,cm$^{-2}$       & $0.3$ frozen        \\
 $\Gamma_1$     & $-$                        & $2.137\,\pm\,0.002$ \\
 $E_{Break}$    & keV                        & $43.0\,\pm\,9.0$    \\
 $\Gamma_2$     & $-$                        & $2.19\,\pm\,0.01$   \\
 norm           & ph\,cm$^{-2}$\,s$^{-1}$\,keV$^{-1}$\footnote{At 1\,keV}            & $9.44\,\pm\,0.03$   \\

 \hline
 \multicolumn{3}{|c|}{Data Group: NuSTAR FPMA}                    \\
 \hline
 Const.         & $-$                        & $1.0$ frozen       \\
 \hline
 \multicolumn{3}{|c|}{Data Group: NuSTAR FPMB}                    \\
 \hline
 Const.         & $-$                        & $0.966\,\pm\,0.002$\\
 \hline
 \multicolumn{3}{|c|}{Data Group: \textit{Swift}-BAT}             \\
 \hline
 Const.         & $-$                        & $1.22\,\pm\,0.27$  \\
 \hline
 \multicolumn{3}{|c|}{Data Group: COSI-Balloon}                   \\
 \hline
 Const.         & $-$                        & $1.022\,\pm\,0.006$\\
 \hline 
 \hline
 $\chi^2/dof$   & $-$                         &503/477            \\
 \hline
\end{tabular}
\end{table}


\subsection{Imaging}\label{sec:imaging}

The imaging response described in Sec.\,\ref{sec:image_resp} is applied to the three data sets. We use this response to perform an image reconstruction using the Richardson-Lucy (RL) deconvolution technique \citep{Richardson1972_RichardsonLucy,Lucy1974_RichardsonLucy}. 
This algorithm has been successfully used in MeV $\gamma$-ray astrophysics \citep[\textit{e.g.}][]{Knoedlseder2005_RL511,Knoedlseder1996_26Al,Knoedlseder1996_COMPTELimaging}, and can produce less-biased images of sources with unknown locations.
The RL algorithm has been proven to converge to the maximum likelihood solution \citep{Shepp1982_RL} but also has the tendency to amplify noise peaks in the heavily background-dominated data of MeV instruments \citep{Knoedlseder1999_26AlCOMPTEL}.
The algorithm is described by the iterative update of an initial map by forward and backward application of the image response such that:
\begin{equation}
M_j^{k+1} = M_j^k + \delta M_j^k \\
\label{equ:RL}
\end{equation}
\begin{equation}
\delta M_j^k = \displaystyle \frac{M_j^k}{\displaystyle \sum_{i} R_{ij}} \sum_{i} \left(\frac{d_{i}}{m_{i}}-1\right) R_{ij}
\label{equ:Mstep}
\end{equation}

\noindent where $M_j^k$ is the k-th image proposal (`map,' with image space indexed by $j$) and $R_{ij}$, is the instrument imaging response and iteratively updated $\delta M_j^k$.
The application of the imaging space into data space would be forward folding (Eq.\,\ref{equ:model}), whereas the application from data space to image space would be similar to a backward projection of data space counts onto the celestial sphere (Eq.\,\ref{equ:Mstep}). 
This also includes background photons so that a single back projection might identify hot spots in the image dimension.
\par
Weakly exposed regions in the data sets of Poisson-count-limited experiments like COSI-Balloon are prone to artifacts, as individual fluctuations in a sparsely populated data space can lead to false high-significance regions in the reconstructed image. 
This effect is common near the edge of exposure where the signal-to-noise is particularly high.
For this reason, we apply a lower limit to the total exposure in each bin that is included in the RL image, effectively removing these bins along the edge of exposure.
The exposure maps for all three sources are shown in Figure\,\ref{fig:expo_maps}, where the lower limit of exposure described above is represented by the red tracer.
Furthermore, to reduce bias in the final image, we slightly randomize the expected background amplitudes by 1--10\% for each background time node so that the initial isotropic map used in the RL algorithm is not precisely tuned to the known background count rate.
To determine which iteration of the RL algorithm best represents the data, we follow a similar procedure as outlined in \cite{Siegert2020:511}.

\section{Results}
\subsection{Crab} \label{sec:Crab_results}

The image shown in Figure\,\ref{fig:CRAB_Swift_BAT_spec} (left) was generated using the Richardson-Lucy deconvolution technique applied to the observation of the Crab nebula and internal pulsar in the 325\,--\,480\,keV band.
This energy band is where the COSI-Balloon instrument reaches its peak efficiency.
After 300 iterations of the RL algorithm, the Crab is clearly visible as the brightest pixel in the image, and this serves as a demonstration of the achieved angular resolution of COSI-Balloon (${\sim}8^{\circ}$, 325\,--\,480\,keV).
Using a model-fitting approach that estimates the total background photon and source photon counts, the Crab was detected at a significance of ${\sim}7\sigma$ in this energy band and is detected at $11\,\sigma$ across the full COSI-Balloon energy band (220\,keV\,--\, 5000\,keV).

\par
Given that the Crab is located in the Galactic anti-center, the best observations of the source would be from the northern hemisphere.
As shown in Figure\,\ref{fig:pointings}, during the 2016 balloon flight the Crab was only visible far off-axis in the instrument FOV once the flight path reached its northernmost latitude, resulting in a total observation time of $T_{obs} = 0.55$\,Ms.
The Crab $\gamma$-ray emission observed by COSI-balloon passed through a significant amount of atmosphere where the scattering of the $\gamma$-rays occurs.
These scattered photons can still enter the detector volume but at different angles and with lower energies, which can result in the flux enhancements seen around the brightest pixel at the known location of the Crab (Figure\,\ref{fig:CRAB_Swift_BAT_spec}, left) \citep{Karwin:2024kcl}.
The reduction of the photo-peak flux due to atmospheric absorption is included in the response used to generate this image.
All other features in the background of this image are consistent with noise from background fluctuations.

\par
The flux measured for the central pixel and the eight neighboring pixels in the 325\,--\,480\,keV energy band is (4.5\,$\pm$\,1.6)\,$\times\,10^{−3}$\,ph\,cm$^{-2}$\,s$^{-1}$.
This flux measurement is within ${\sim}\,8\%$ of the expected value of $4.1\,\times\,10^{−3}$\,ph\,cm$^{-2}$\,s$^{-1}$ (\cite{Jourdain2020_Crab}, Table 2, 2012–2019 values in this energy band).
Accounting for the full bandwidth of the instrument, the total flux measured during the flight is (1.8\,$\pm$\,0.7)\,$\times\,10^{−2}$\,ph\,cm$^{-2}$\,s$^{-1}$. 
This accounts for the 8$^{\circ}$ angular resolution of COSI-Balloon in this energy range with 6$^{\circ}$ pixels.
%

\begin{figure}[!t]
    \centering
    \includegraphics[width=0.45\textwidth,trim=0.0in 0.0in 0.0in 0.0in,clip=True]{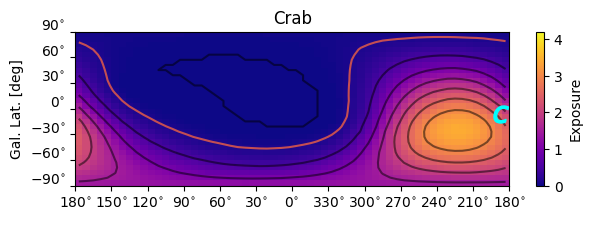}
    \includegraphics[width=0.45\textwidth,trim=0.0in 0.0in 0.0in 0.0in,clip=True]{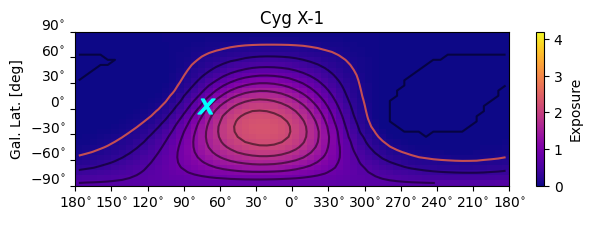}
    \includegraphics[width=0.45\textwidth,trim=0.0in 0.0in 0.0in 0.0in,clip=True]{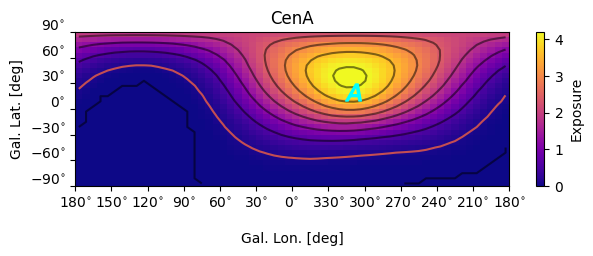}
    \caption{\footnotesize Exposure maps of the three sources during the 2016 balloon flight. The cyan letters $C$, $X$, and $A$ represent the known source locations of Crab, Cyg\,X-1, and Cen\,A, respectively. The black contours represent the different thresholds of exposure, and the red contour represents the minimum exposure limit for pixels to be included in the Richardson-Lucy image. The color map shows the instrument exposure in units of $10^8$\,s\,cm$^2$\,sr$^{-1}$.}
    \label{fig:expo_maps}
\end{figure}

\begin{figure}[!t]
    \centering
    \includegraphics[width=0.49\textwidth,trim=0.0in 0.0in 0.0in 0.0in,clip=True]{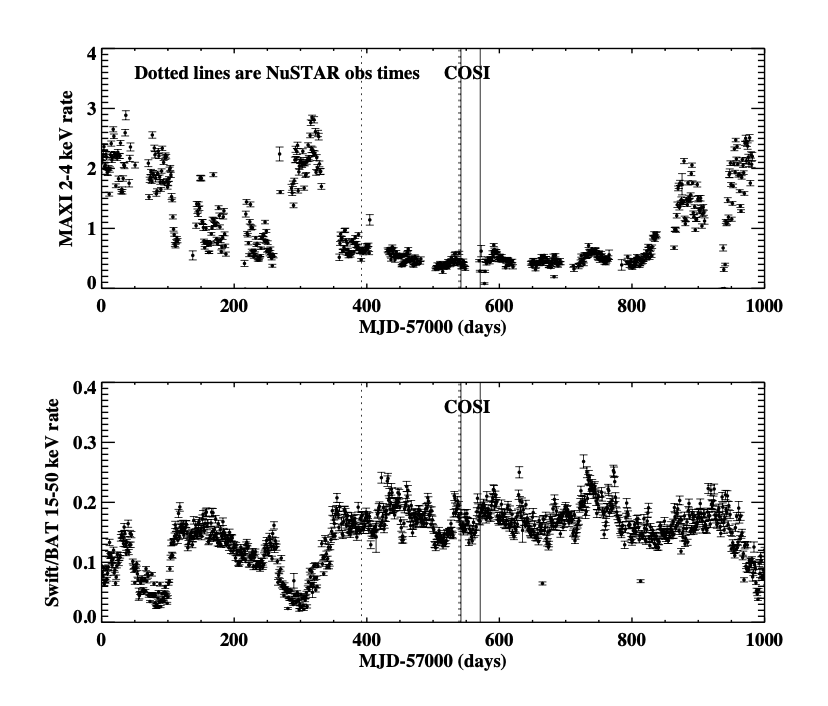}
    \caption{\footnotesize Cygnus\,X-1 flux measured by MAXI and \textit{Swift}-BAT during the COSI-Balloon 2016 flight. The COSI-Balloon flight period is marked by solid vertical lines, and the two NuSTAR observations considered for this analysis in the combined spectrum are marked by two vertical dashed lines. The low 2\,--4\,keV count rate measured by MAXI and the high 15\,--50\,keV count rate measured by \textit{Swift}-BAT are strong indicators that Cyg\,X-1 was in the Low Hard Sate (LHS) during the COSI-Balloon observations. }
    \label{fig:cygx1_state}
\end{figure}

\par
Figure\,\ref{fig:CRAB_Swift_BAT_spec} (right) also shows the combined spectrum of the Crab with observations from NuSTAR (FPMA in black and FPMB in red), \textit{Swift}-BAT (green), and COSI-Balloon (blue). 
The most proximal NuSTAR observation in time of the Crab was Observation ID 10202001002 from May 7\textsuperscript{th}, 2016, with an exposure time of 1.4 ks.
Since the Crab is an almost consistent source in soft $\gamma$-rays, the \textit{Swift}-BAT data set we used is from the 105-month all-sky hard X-ray survey \citep{SWIFT_BAT_105_Month_Survey}. 
The model that best fits the data is a broken power law (bknpower) with an interstellar absorption factor (TBabs) and the fit results are shown in Tab.\,\ref{tab:Crab}.
%


\begin{figure*}[!t]
    \centering
    \includegraphics[width=0.52\textwidth,trim=0.0in 0.0in 0.0in 0.0in,clip=True]{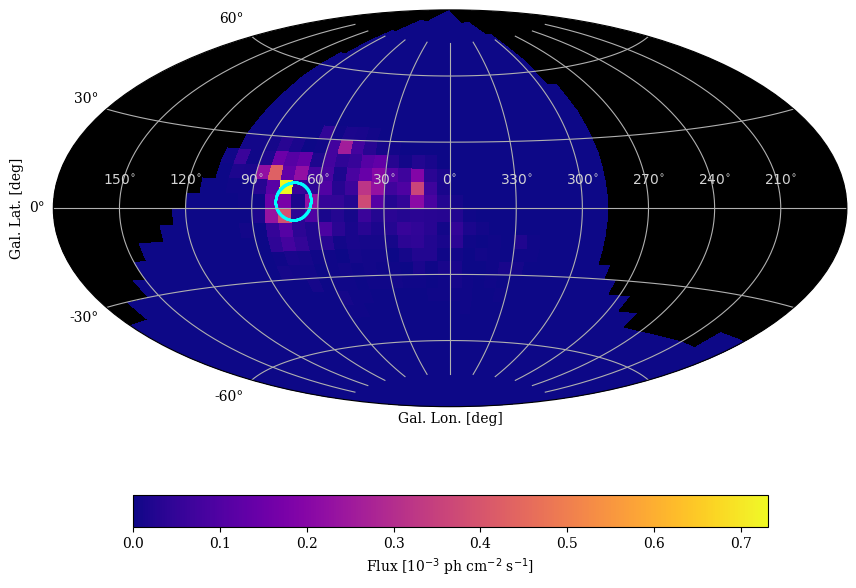}
    \includegraphics[width=0.47\textwidth,trim=0.0in 0.0in 0.0in 0.0in,clip=True]{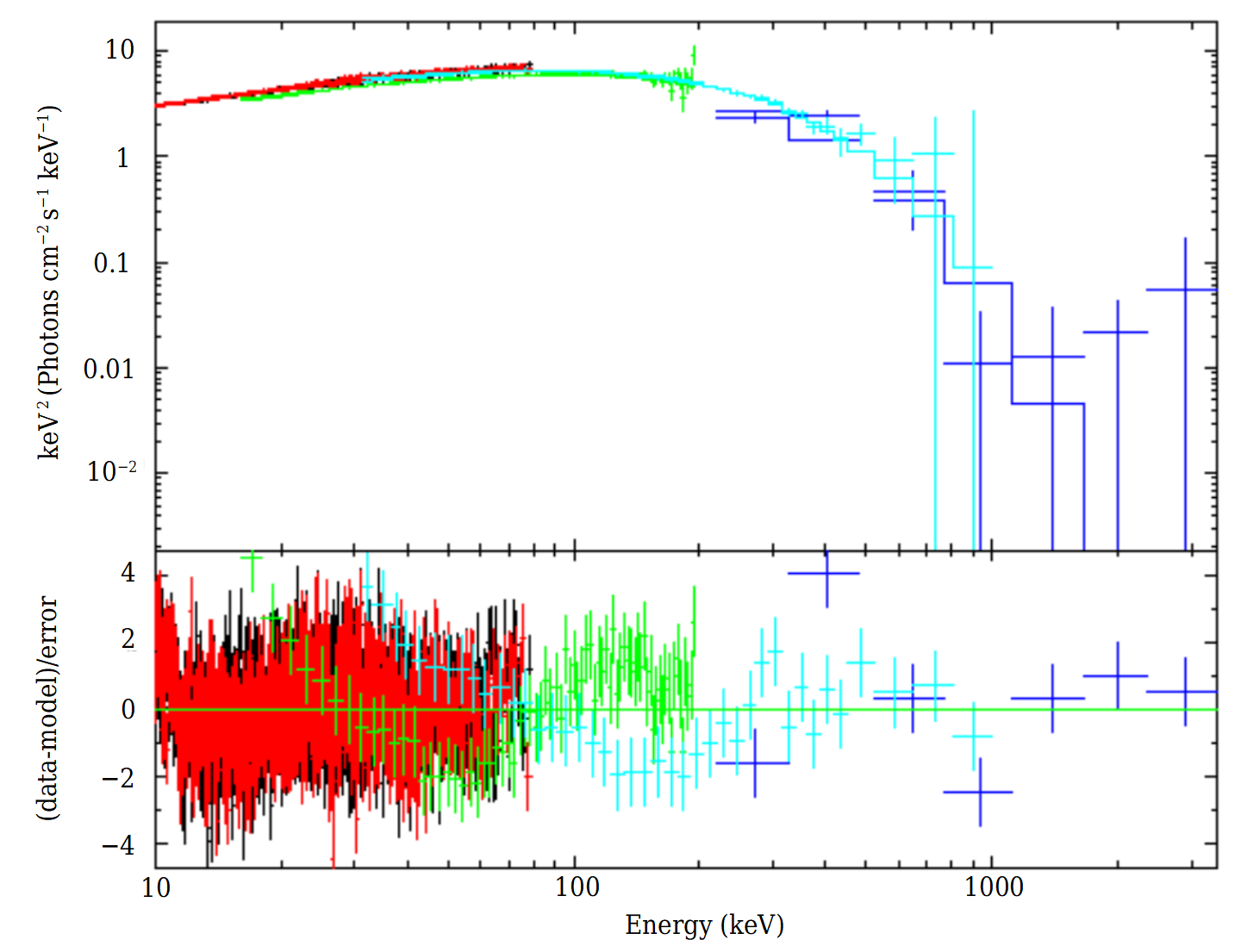}
    \caption{\footnotesize \textbf{Left:} Cygnus\,X-1 RL image generated from the 2016 COSI-Balloon flight data. The pixels are separated into $6^{\circ}$ bins, and the cyan circle drawn at the known source location with the ARM radius of 8\,$^{\circ}$ is shown to represent the instrument resolution for the energy bin 325\,--\,480\,keV  \textbf{Right:} Cygnus X-1 spectrum as
    measured by NuSTAR (black, red), \textit{Swift}-BAT (green), INTEGRAL/IBIS (cyan), and the COSI-Balloon instrument (blue), fit with a cutoff power law.
    The fit has a cutoff energy of 134.8\,keV and reduced-$\chi^2$ value of 1.4 (see Tab.\,\ref{tab:cygx1}).}
    \label{fig:cyg_spec}
\end{figure*}


The known hydrogen column density in the direction of the Crab is $N_H = (3.0\,\pm\,0.5) \times 10^{21}\,$cm$^{−2}$ \citep{Sollerman_2000}; This value was used for the interstellar absorption factor. 
We've also applied atmospheric correction factors to each energy bin to account for atmospheric scattering based on a simplified rectangular atmospheric model.
These atmospheric correction factors and how they have been modeled are described in \cite{Karwin:2024kcl}.
The model fit results in a break in the spectrum at $43\,\pm\,9$\,keV, a photon index below the break of $\Gamma = 2.137\,\pm\,0.002$, and a photon index above the break of $\Gamma = 2.19\,\pm\,0.02$. 
The reduced-$\chi^2$ value for the fit is 1.04 with 482 degrees of freedom. 
This is within 8\% of the photon index reported by INTEGRAL/SPI, $\Gamma = 2.32\,\pm\,0.02$ \citep{Jourdain2020_Crab}.
In \cite{Churazov2007_CXB_INTEGRAL}, the absolute normalization of the Crab spectrum between different hard X-ray instruments were found to be as high as 30\,\%; The normalization is directly related to the photon index of the measurements. 
Therefore, it is understandable that the COSI-Balloon prototype measurement in the MeV bandpass is a few percent off when compared to what has been reported by INTEGRAL.
We also fit the spectra with a power law, broken power law, and power law with pegged normalization models.
The broken power law model best represented the data set that included the COSI-Balloon spectral fits.

\subsection{Cygnus\,X-1} \label{sec:CygX1_results}

Figure\,\ref{fig:cyg_spec} (left) shows the image generated by the Richardson-Lucy deconvolution technique applied to the observations of Cyg\,X-1; also in the 325\,--\,480\,keV band.
After 300 iterations of the RL algorithm, Cyg\,X-1 is visible as the brightest pixel in the image, and the pixel falls within the $8^{\circ}$ angular resolution measure from the known source location. 
The angular resolution of the image is also limited by $6^{\circ}$ pixels.
Cyg\,X-1 was detected at a significance of 6\,$\sigma$ within the energy range 325\,--\,480\,keV bandwidth a significance of 9\,$\sigma$ for the full COSI band (220\,keV\,--\,5000\,MeV).
As shown in Figure\,\ref{fig:pointings}, during the 2016 balloon flight, Cyg\,X-1 was visible at higher angles from instrument zenith once the flight path reached its most Northern latitudes.
The exposure of Cyg\,X-1 is shown in  Figure\,\ref{fig:expo_maps} for this data set with the exposures of the Crab and Cen\,A also included for comparison.
Cyg\,X-1 was often on the edge of exposure throughout the flight.
The $\gamma$-ray  emissions from Cyg\,X-1, therefore, also passed through a significant amount of atmosphere where photon scattering occurs.
The combination of less exposure to the Cyg\,X-1 region ($T_{obs} = 0.35$\,Ms) and the observations being restricted to higher incident angles resulted in a limited data set that can produce flux enhancements in the image in pixels around the known location of Cyg\,X-1 (Figure\,\ref{fig:cyg_spec}, left).
As RL deconvolution has a tendency to magnify the background flux of pixels with very low statistics, the red contour shown in Figure\,\ref{fig:expo_maps} represents the minimum-allowed exposure of a pixel included in the RL image.

\par
The reduction of the photo-peak flux due to atmospheric absorption is also included in the response used to 
generate this image.
The measured flux of Cyg\,X-1 for the central pixel and the eight neighboring pixels is (1.4\,$\pm$\,0.2)\,$\times\,10^{−3}$\,ph\,cm$^{-2}$\,s$^{-1}$.
The total flux measured for the full COSI-Balloon energy band is (5.8\,$\pm$\,0.8)\,$\times\,10^{−3}$\,ph\,cm$^{-2}$\,s$^{-1}$.
For comparison, COSI-Balloon's flux measurement within the 200\,--\,1000\,keV energy band ($3.9\,\times\,10^{-9}$\,ergs\,cm$^{-2}$\,s$^{-1}$, $8.2\,\times\,10^{−3}$\,ph\,cm$^{-2}$\,s$^{-1}$) with Cyg\,X-1 in the LHS is within ${\sim}10\%$ of the measured flux reported by INTEGRAL ($3.5\,\times\,10^{-9}$\,ergs\,cm$^{-2}$\,s$^{-1}$, \cite{Cangemi2021_CygX1_hardtail}). 
\\

Figure\,\ref{fig:cyg_spec} (right) shows the combined spectrum of Cyg\,X-1 with observations from NuSTAR (FPMA in black and FPMB in red), \textit{Swift}-BAT (green), INTEGRAL/IBIS (cyan), and COSI-Balloon (blue). 
Two observations of Cyg\,X-1 performed by NuSTAR occurred while the source was in the hard state and proximal in time to the COSI-Balloon flight: Observation ID 90101019002 (Jan. 5\textsuperscript{th}-- 6\textsuperscript{th} 2016), ID 30002150006 (May 31\textsuperscript{st} 2016).
These two observations are shown in Figure\,\ref{fig:cygx1_state} as vertical dashed lines with the COSI-Balloon observation of Cyg\,X-1 represented by solid vertical lines.
The latter NuSTAR observation took place during the COSI-Balloon flight, and we used this observation in our spectral analysis.
Cyg\,X-1 was observed by COSI-Balloon from MJD 57542.0 to 57571.0 (06-03-2016 through 07-02-2016). 
\textit{Swift}-BAT data of Cyg\,X-1 was processed for this observation period using BATIMAGER \citep{BATIMAGER_Segreto_2010} and is used in this analysis. 
Atmospheric effects were accounted for in the COSI-Balloon data via the application of atmospheric scattering correction factors to the COSI-Balloon spectral fits using the average zenith angle of the source during observations \citep{Karwin:2024kcl}.
The best model to fit the combined spectra that includes the COSI-Balloon spectral fits is a cutoff power law.
%
The fit results (Tab.\,\ref{tab:cygx1}) indicate a cutoff energy of $138.3\,\pm\,1.0$\,keV and a photon index of $\Gamma = 1.358\,\pm\,0.002$.
The reduced\,-\,$\chi^2$ value of 1.32 with 1686 degrees of freedom.


\begin{table}[t]
\centering
\caption{Spectral Results for Cyg\,X-1 Cutoff Power Law Fit\label{tab:cygx1}}
\begin{tabular}{| c | c | c | }
\hline
\hline
 Parameter\footnote{The errors on the parameters are 90\% confidence.}       &Unit            &Value\\ 
 \hline
 \hline
 \multicolumn{3}{|c|}{Tied Parameters\footnote{The full model in XSPEC is: constant*cutoffpl}} \\
 \hline
 $\Gamma$       & $-$            & $1.358\,\pm\,0.002$  \\
 $E_{cutoff}$   & keV             & $138.3\,\pm\,0.9$   \\
 norm           & ph\,cm$^{-2}$\,s$^{-1}$\,keV$^{-1}$\footnote{At 1\,keV}            & $0.731\,\pm\,0.003$   \\

 \hline
 \multicolumn{3}{|c|}{Data Group: NuSTAR FPMA}          \\
 \hline
 Const.         & $-$            & $1.0$ frozen         \\
 
 \hline
 \multicolumn{3}{|c|}{Data Group: NuSTAR FPMB}          \\
 \hline
 Const.         & $-$            & $1.0112\,\pm\,0.0008$\\

 \hline
 \multicolumn{3}{|c|}{Data Group: \textit{Swift}-BAT}   \\
 \hline
 Const.         & $-$            & $0.876\,\pm\,0.002$  \\

 \hline
 \multicolumn{3}{|c|}{Data Group: COSI-Balloon}         \\
 \hline
 Const.         & $-$            & $0.71\,\pm\,0.06$    \\

 \hline 
 \multicolumn{3}{|c|}{Data Group: INTEGRAL/IBIS}        \\
 \hline
 Const.         & $-$            & $0.957\,\pm\,0.005$  \\

 \hline
 \hline
 $\chi^2/dof$   & $-$             &2236/1693            \\
 \hline
\end{tabular}
\end{table}


\subsection{Centaurus A} \label{sec:CenA_results}

\begin{figure*}[!t]
    \centering
    \includegraphics[width=0.52\textwidth,trim=0.0in 0.0in 0.0in 0.0in,clip=True]{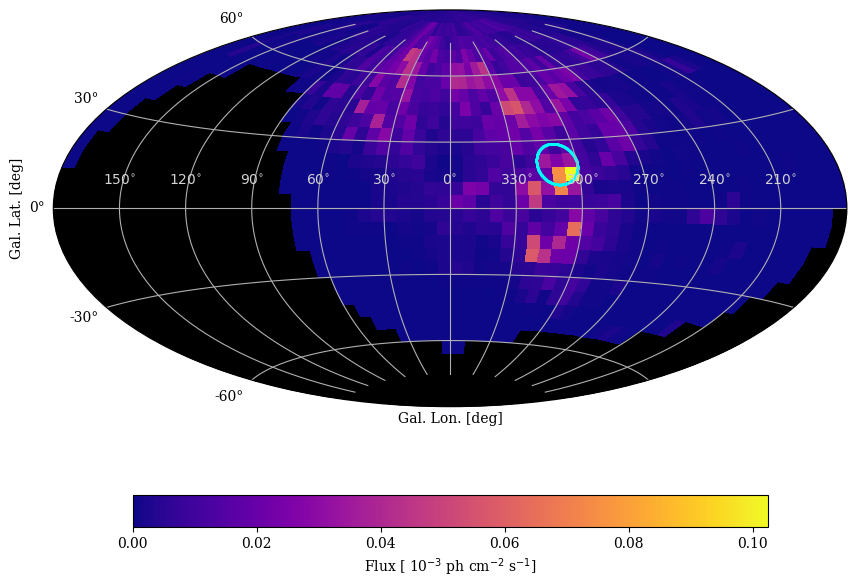}
    \includegraphics[width=0.47\textwidth,trim=0.0in 0.0in 0.0in 0.0in,clip=True]{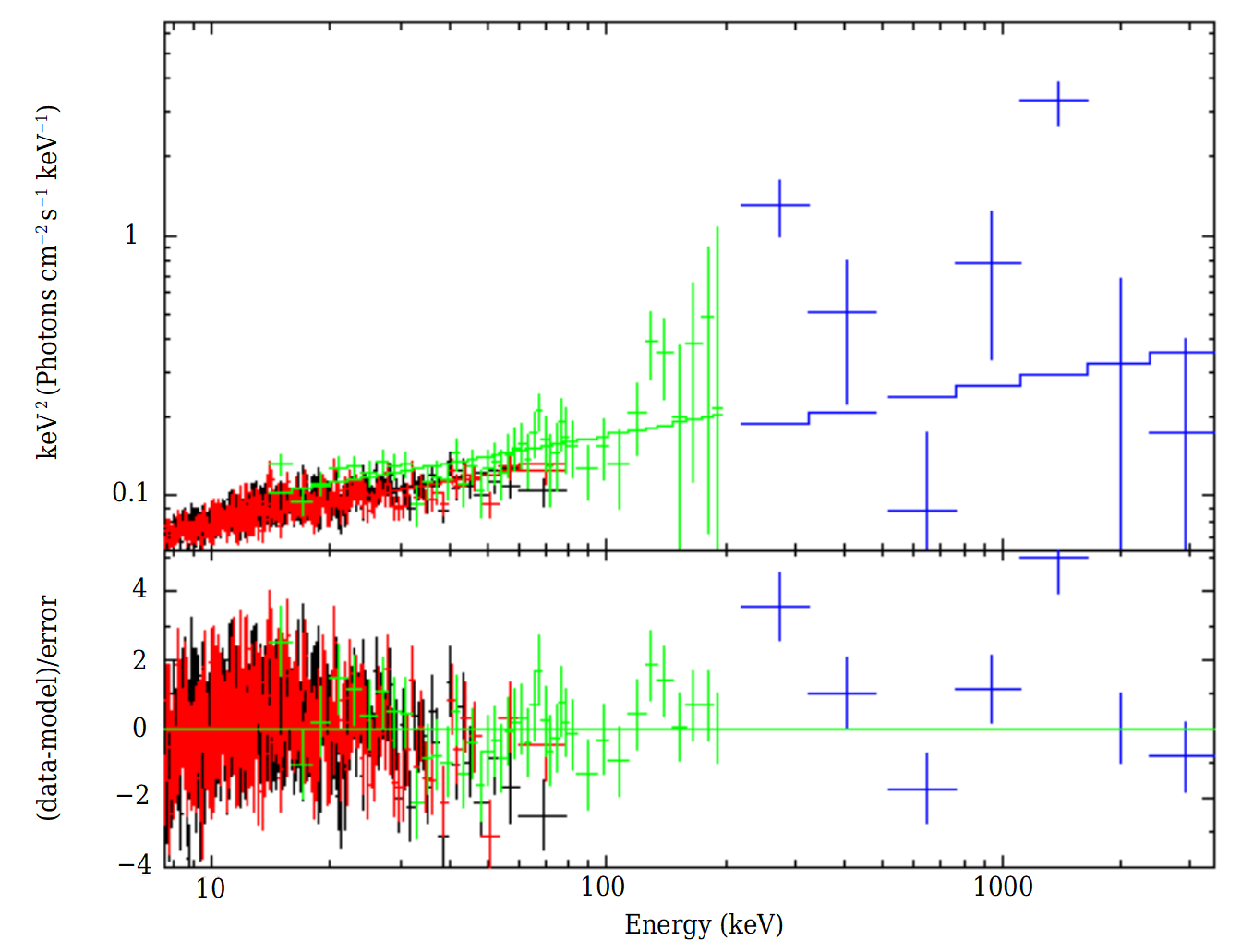}
    \caption{\footnotesize \textbf{Left:} Centaurus\,A image generated from the 2016 COSI balloon flight data. The cyan circle drawn at the known source location with the ARM radius of 9\,$^{\circ}$ is shown to represent the instrument resolution for the energy bin 220\,--\,325\,keV. \textbf{Right:} Centaurus A spectrum as measured by NuSTAR (black, red), \textit{Swift}-BAT (green), and the COSI-Balloon instrument (blue), fit with a power law. The fit has a photon index of $\Gamma$ = 1.732\,$\pm$\,0.008 and a reduced-$\chi^2$ value of 1.2 (see Tab.\,\ref{tab:cena}).}
    \label{fig:cena_spec}
\end{figure*}

Figure\,\ref{fig:cena_spec} (left) shows the image generated by the Richardson-Lucy deconvolution technique applied to the observations of Cen\,A in the 220\,--\,325\,keV band.
After 300 iterations, the region of Cen\,A shows a bright pixel within $9^{\circ}$ of the known source location. 
Cen\,A was detected at a significance of 9\,$\sigma$ within the energy range 220\,--\,325\,keV and a significance of 13\,$\sigma$ for the full COSI band.\footnote{The diffuse emission along the Galactic plane and the bulge emission at 511\,keV may result in inflated significance calculations for the COSI-Balloon measurement of Cen\,A.}

As shown in Figure\,\ref{fig:pointings}, during the 2016 balloon flight, Cen\,A was visible at higher off-axis angles in the instrument's FOV for long periods during observations for a total observation time of $T_{obs} = 0.7$\,Ms.
Due to high background rates uncorrelated with altitude in the early flight, we reject all data during this period.
This may have impacted the quality of this data set, being that Cen\,A was observed much higher in the instrument FOV during this portion of the flight.
A $60^{\circ}$ pointing cut was applied to the image data selection to better isolate the source from diffuse emission from the Galactic plane.
COSI-Balloon measured a flux of (0.2\,$\pm$\,0.1)\,$\times\,10^{−3}$\,ph\,cm$^{-2}$\,s$^{-1}$ of Cen\,A within the 220\,--\,325\,keV band and (1.0\,$\pm$\,0.4)\,$\times\,10^{−3}$\,ph\,cm$^{-2}$\,s$^{-1}$ for the full COSI energy band.
For comparison, COMPTEL reported a time-averaged flux of Cen\,A in the 0.75\,--1.00\,MeV band of $0.091$\,$\pm$\,0.03\,$\times\,10^{−3}$\,ph\,cm$^{-2}$\,s$^{-1}$ \citep[Table 2, ][]{cCOMPTEL_CenA}.

The spectra shown in Figure\,\ref{fig:cena_spec} (right) combine observations from NuSTAR (FPMA in black and FPMB in red), \textit{Swift}-BAT (green), and COSI-balloon (blue).
Data from NuSTAR observation ID 60101063002 from May 17\textsuperscript{th} 2015 with 22.5 ks of exposure was used because it was the closest observation of the source in time. 
The \textit{Swift}-BAT data points were gleaned from the 105-month all-sky hard X-ray survey \citep{SWIFT_BAT_105_Month_Survey}.
Because the Crab was in the FOV during parts of Cen\,A observations, simulated Crab emissions were included in the background model for the Cen\,A spectral fits.
Similar to the Crab analysis, atmospheric scattering correction factors were applied to the Cen\,A spectral fits \citep{Karwin:2024kcl}.
The model that best fits the combined spectra (Tab.\,\ref{tab:cena}) that includes COSI-Balloon spectral fits is a power law with a photon index of $\Gamma = 1.732\,\pm\,0.008$. 
The reduced-$\chi^2$ value for this model fit is $1.2$ with 608 degrees of freedom.
Due to spatial correlations, we estimate the systematic uncertainty of the Cen\,A flux above 500\,keV to be 1 order of magnitude, and below 500\,keV to be closer to 30\,\%.
Compared to the results presented by \cite{Rodi_2023}, where NuSTAR, INTEGRAL/SPI, and INTEGRAL/ISGRI, were fit with a power law that includes a redshift component (\texttt{zpowerlw}) a photon index of 1.807\,$\pm$\,0.002 was reported.
\\

\section{Summary and Discussion}
We present the images and spectral models of three $\gamma$-ray sources (Crab, Cyg\,X-1, and Cen\,A) as measured by the COSI-Balloon instrument from data obtained during a SPB flight in 2016.
The variability of the background environment for a balloon payload is presented and we describe our method to perform data selections to optimize the data set, generate accurate instrument and background responses, and model the background throughout the varying flight conditions.
\par
The Crab nebula and pulsar were imaged using the Richardson-Lucy image deconvolution technique and the known location of the Crab is revealed in this image, demonstrating COSI-Balloon's ability to measure a luminous source within its instrumental bandwidth and angular resolution. 
The flux measured in the 325\,--\,480\,keV energy band is (4.5\,$\pm$\,1.6)\,$\times\,10^{−3}$\,ph\,cm$^{-2}$\,s$^{-1}$.
Combined spectra, including observations from NuSTAR, \textit{Swift}-BAT, and COSI-Balloon, produced a best-fit model to a broken power law with a photon index above the break of $\Gamma = 2.19\,\pm\,0.02$ which is in good agreement with previous measurements \citep[$\Gamma = 2.32\,\pm\,0.02$][]{Jourdain2020_Crab}. 

\par
The image of Cyg\,X-1 is presented where the brightest pixel is within the $8^{\circ}$ angular resolution measure of the COSI-Balloon instrument, demonstrating the ability to produce an image of a $\gamma$-ray source that is ${\sim}3\,\times$ less luminous than the Crab and with significantly less exposure.
The model that best fit the combined spectra that includes observation from NuSTAR, \textit{Swift}-BAT, INTEGRAL/IBIS, and COSI-Balloon, was a cutoff power law with a cutoff energy of $138.3\,\pm\,1.0$\,keV and a photon index of $\Gamma = 1.358\,\pm\,0.002$.
This cutoff energy of Cyg\,X-1 while in the LHS is in good agreement with previous estimates of a high energy cutoff ${\sim}100$\,keV.
It may also provide further evidence of the existence of compact jet emissions extending into MeV energies while this X-ray binary source is in the LHS \citep{Fender_cygx1_jet_2001,Sterling_cygx1_jet_2001}.

\par
The radio-loud AGN, Cen\,A, image is presented with the brightest pixel falling within the ${\sim}9^{\circ}$ ARM of the COSI instrument in the 220\,--\,325\,keV bandpass.
This image has significant background artifacts that can be attributed to the quality of the exposure to the source, its low luminosity (${\sim}20\,\times$ less luminous than the Crab), and possible wrong flux assignments from Galactic diffuse emission.
A power law model best fits the combined spectra that included NuSTAR, \textit{Swift}-BAT, and COSI-Balloon spectral fits, with a photon index of $\Gamma = 1.732\,\pm\,0.008$.
There is considerable interest within the high-energy astrophysics community to investigate the existence of a hard-tail extending above 1\,MeV, an existence possibly corroborated by the COSI-Balloon data.
Given the low luminosity of Cen\,A and the systematic uncertainties of COSI-Balloon, it is unlikely that this data set can further elucidate the potential presence of a hard-tail in the soft $\gamma$-ray Cen\,A spectrum.

\par
We note that the image reconstruction and flux extraction in the Cen\,A dataset results in larger systematic uncertainties.
This is seen by the more prominent image artefacts in Figure\,\ref{fig:cena_spec} (left) and the jumpy spectral data points for Cen\,A.
Especially in this dataset, the diffuse emission \citep{Karwin_2023} along the Galactic plane, strong gamma-ray sources, and especially the galactic positron annihilation emissions, \citep{Kierans2019_511COSI, Siegert2020:511} may lead to the erroneous assignment of flux values by using the Richardson-Lucy algorithm without regularization. 
This will be improved in upcoming publications using the full instrument response in \texttt{cosipy} (Yoneda \textit{et al.} 2024, in prep.).

\par
The \textit{MeV Gap} is a relatively unexplored region of the electromagnetic spectrum that is ripe with potential for key, impactful discoveries in astrophysics.
Measurements of the Crab, Cygnus\,X-1, and Centaurus\,A with the 2016 COSI-balloon instrument are key milestones in our analysis methods, demonstrating the ability to create response matrices that produce source fluxes and spectral parameters consistent with measurements from other instruments, the successful development of a variable background model, and improved data reduction.
These measurements also serve as blueprints for future studies of $\gamma$-ray sources with the COSI satellite mission at MeV energies.
The single photon reconstruction with strong background suppression makes the COSI instrument design ideal for  $\gamma$-ray spectroscopy and imaging studies.
The COSI satellite in its low-Earth equatorial orbit will benefit from a low, stable background relative to the balloon instrument and will not have to contend with the complications of atmospheric attenuation.
Measurement of these sources will be significantly improved when performed from a favorable space environment \citep{Cumani_2019}. 
\par

The COSI satellite instrument will also benefit from an additional layer of four germanium detectors that will increase the effective area, a reduced strip-pitch of 1.162\,mm (vs. 2\,mm) that will improve the instrument's angular resolution, and an improved bismuth germanium oxide (BGO) active shield design that will further reduce instrumental background. 
In addition, the COSI science team is developing advanced analysis techniques and software tools via annual Data Challenge software releases \citep{COSI_Data_Challenge_2022HEAD} with the aim of optimizing the analysis pipeline, \texttt{COSIpy}, for the upcoming satellite mission \citep{Martinez_2023_COSIpy}.
Given all of these improvements over the balloon instrument and environment, the COSI Small Explorer mission will be a powerful observatory for the study of the emission of MeV $\gamma$-ray sources. 


\begin{table}[t]
\centering
\caption{Spectral Results for Cen\,A Power Law Fit\label{tab:cena}}
\begin{tabular}{| c | c | c | }
\hline
\hline
 Parameter\footnote{The errors on the parameters are 90\% confidence.}       &Unit            &Value\\ 
 \hline
 \hline
 \multicolumn{3}{|c|}{Tied Parameters\footnote{The full XSPEC model is: constant*powerlaw}} \\
 \hline
 $\Gamma$       & $-$            & $1.732\,\pm\,0.008$\\
 norm           & ph\,cm$^{-2}$\,s$^{-1}$\,keV$^{-1}$\footnote{At 1\,keV}          & $0.0432\,\pm\,0.0009$   \\
 
 \hline
 \multicolumn{3}{|c|}{Data Group: NuSTAR FPMA}        \\
 \hline
 Const.         & $-$            & $1.0$              \\
 
 \hline
 \multicolumn{3}{|c|}{Data Group: NuSTAR FPMB}        \\
 \hline
 Const.         & $-$            & $0.981\,\pm\,0.007$\\

 \hline
 \multicolumn{3}{|c|}{Data Group: \textit{Swift}-BAT} \\
 \hline
 Const.         & $-$            & $1.15\,\pm\,0.03$  \\
 \hline
 \multicolumn{3}{|c|}{Data Group: COSI-Balloon}       \\
 \hline
 Const.         & $-$            & $1.0\,\pm\,0.3$    \\
 \hline 
 \hline
 $\chi^2/dof$   & $-$             &759/613            \\
 \hline
\end{tabular}
\end{table}


\section{Acknowledgments} \label{sec:ackowledgements}
The COSI-Balloon program was supported through NASA-APRA grants NNX14AC81G and 80NSSC19K1389. 
We also acknowledge the support for this work under the NASA-APRA grant 80NSSC21K815.
Hiroki Yoneda is supported by the Bundesministerium f\"{u}r Wirtschaft und Energie via the Deutsches Zentrum f\"{u}r Luft- und Raumfahrt (DLR) under contract number 50 OO 2219.
Chris Karwin's research was supported by an appointment to the NASA Postdoctoral Program at NASA Goddard Space Flight Center, administered by Oak Ridge Associated Universities under contract with NASA.
Israel Martinez Castellanos' research is based upon work supported by NASA under award number 80GSFC21M0002.
This work is also supported in part by CNES.
We thank NASA's Balloon Program Office and Columbia Scientific Balloon Facility for their support of the COSI-Balloon SPB campaigns.
We want to express a particular appreciation to team members from NuSTAR, \textit{Swift}-BAT, and INTEGRAL collaborations for their support in preparing and sharing data products used in this analysis.

\bibliography{Roberts_COSI2016}{}
\bibliographystyle{aasjournal}

\end{document}